\documentclass[twoside,fleqn]{article}
\usepackage{espcrc2}
\usepackage{amstex}
\usepackage{dcolumn}

\usepackage{graphicx}

\newcommand{\psibar}{\bar\psi}
\newcommand{\chibar}{\bar\chi}
\newcommand{\U}[1]{\mathrm{U}(#1)}

\newcommand{\One}{\Bbb I}
\newcommand{\cc}[1]{\multicolumn{1}{c}{#1}}
\newcolumntype{d}{D{.}{.}{-1}}
\newlength{\colw}
\title{Numerical study of dense adjoint 2-color matter
\thanks{Material based on talks by L.~Scorzato and 
J.~Skullerud}}

\author{Simon Hands\address[Swansea]{Department of Physics, University
        of Wales Swansea, Singleton Park, Swansea SA2 8PP, Wales},
	Istv\'an Montvay\address[DESY]{Theory Division, DESY,
        Notkestra{\ss}e 85, D-22603 Hamburg, Germany},
	Manfred Oevers\address[Glesca]{Department of Physics and
        Astronomy, University of Glasgow, Glasgow G12 8QQ, Scotland}
	\thanks{Current address: IBM UK, 1 New Square, Bedfont Lakes, 
	Feltham, Middx, TW14 8HB, UK},
	Luigi Scorzato\addressmark[Swansea],
	Jonivar Skullerud\addressmark[DESY]}

\begin{document}
\makeatletter
\@mathmargin = 0pt
\makeatother

\begin{abstract}
We study the global symmetries of SU(2) gauge theory with $N$ flavors
of staggered fermions in the presence of a chemical potential.  We
motivate the special interest of the case $N=1$ (staggered) with
fermions in the adjoint representation of the gauge group.  We present
results from numerical simulations with both hybrid Monte Carlo and
the Two-Step Multi-Bosonic algorithm.
\vspace{1pc}
\end{abstract}

\maketitle

\section{INTRODUCTION}
\label{sec:intro}

QCD at finite baryonic density is expected to display a rich phase
diagram \cite{ARW}. Unfortunately numerical simulations at non zero density
are still prohibitive because of the sign problem (for a review of the recent
progress on that matter see \cite{Cha}). The finite density sector
can be studied numerically in a class of theories
which includes any SU(2) gauge theory.
In such cases, in fact,
the fermionic determinant is real (and positive for an even number
of flavors). All these theories can also be studied by means of 
Chiral Perturbation Theory ($\chi$PT) even in the finite density region
\cite{KSTVZ}.
Here $\chi$PT predicts the presence of some Goldstone modes associated
with spontaneous chiral symmetry breaking which are sensitive to the
chemical potential, and can therefore be called {\em diquark} or {\em baryonic}
Goldstone modes. These modes drive the onset transition between the low
and the high density regime which appears when $\mu$ is of the order
of the pion mass $m_\pi \sim \sqrt{m}$. The same features that ensure that
the fermionic determinant is real are also related to the prediction 
of  an early onset transition. 
It is therefore interesting to note that there is a special case:
the SU(2) gauge theory with one flavor of staggered fermions in the adjoint
representation of the gauge group 
($\text{SU(2)}^{N=1\, \text{stagg}}_{\text{Adj}}$ \cite{HM2}) which
has a real fermionic determinant, but where no baryonic Goldstone modes are
expected.
It is therefore an ideal model to explore the influence of the sign on
the observables and on the pattern of symmetry breaking.  We will give
numerical evidence that this model, if constrained to the sector with
positive determinant, still belongs to the class of theories with
baryonic Goldstone modes. Then we show how the inclusion of the
sign changes this picture, suggesting a delayed onset transition.

As a by-product we gain important experience with two 
numerical algorithms when $\mu\not=0$: 
the hybrid Monte Carlo (HMC) and Two-Step 
Multi-Bosonic algorithms.

\section{SYMMETRIES}
\label{sec:symmetries}

We first present the continuum and lattice formulation of Two Color QCD. 
The fermionic part of the action for $N_f$ flavors $f$ in the
continuum is:
\begin{equation}
S_F = \int\!\!d^4x\,\psibar^f(x) (\partial_\nu\!+\! igA_\nu\! +\!
\mu\delta_{\nu0})\gamma_\nu\psi^f(x)
\label{eq:s-cont}
\end{equation}
The lattice action for staggered fermions is
\begin{equation}
\begin{split}
S_F &=\sum_{x,y}\bar\chi^p(x)\bigl(D_{xy}[U,\mu]+m\delta_{xy}\bigr)\chi^p(y)\\
 &\equiv\sum_{x,y}\bar\chi^p(x)M_{xy}[U,\mu]\chi^p(y)\, ,
\end{split}\label{eq:s0}
\end{equation}
where the index $p$ runs over $N$ flavors of staggered quark, 
and $D$ is given by
\begin{equation}
\begin{split}
D_{xy} &= \sum_{\nu\not=0}\frac{\eta_\nu(x)}{2}
\bigl(U_\nu(x)\delta_{x,y-\hat\nu}
        \!-\!U_\nu^\dagger(y)\delta_{x,y+\hat\nu}\bigr)\\
+& \frac{\eta_0(x)}{2}\bigl(e^\mu U_0(x)\delta_{x,y-\hat0}
        \!-\!e^{-\mu}U_0^\dagger(y)\delta_{x,y+\hat0}\bigr)\label{eq:s1}
\end{split}
\end{equation}
$\chi, \bar\chi$ are single spin component Grassmann objects, 
and the phases $\eta_\mu(x)$ are defined to be $(-1)^{x_0+\cdots+x_{\mu-1}}$.
For brevity, we have here used $U$ to denote both the fundamental
($2\times2$ complex unitary) and adjoint ($3\times3$ real orthogonal)
link fields.

\subsection{Sign of the determinant}
\label{subs:sign}

As is well known, Two Color QCD has a real fermionic determinant also at
non zero density. In fact the SU(2) group only has real or 
pseudoreal representations. For all such representations it is always possible
to find a unitary operator $T$ such that $KT$  commutes with the 
fermionic matrix $M$, where $K$ is the complex conjugate operator. 
If such an operator exists one can show \cite{KSTVZ,us} that
all the eigenvalues of $M$ appear in complex conjugate pairs and
thus the determinant of $M$ must be real. 
Although for the theories we are considering here $\det{M}$ is
real, it need not be positive definite.  Indeed, it turns out that
some of these theories have a sign problem.  It can be shown \cite{KSTVZ,us}
that a sufficient condition for a positive definite action is that
there is one choice of $T$ such that $(KT)^2=-1$.

For the continuum action, given by (\ref{eq:s-cont}), one finds
\begin{equation}
\begin{split}
\text{fundamental:}\;\;\;T&=C\gamma_5\otimes\tau_2,\;\;(KT)^2=1 \,;\\
\text{adjoint:}\;\;\;T&=C\gamma_5\otimes\One,\;\;(KT)^2=-1 \, .
\end{split}
\label{eq:contT}
\end{equation}
Here, $C$ is the charge conjugation matrix.
For staggered lattice fermions $M$ is given by (\ref{eq:s0},\ref{eq:s1}), and
\begin{equation}
\begin{split}
\label{eq:lattT}
\text{fundamental:}\;\;\;T&=\tau_2,\;\;(KT)^2=-1 \, ; \\
\text{adjoint:}\;\;\;T&=\One,\;\;(KT)^2=1 \, .
\end{split}
\end{equation}
We thus have a proof that the
functional integral measure is positive definite for continuum adjoint 
quarks and fundamental staggered quarks. There is no such proof
for continuum fundamental quarks and staggered adjoint quarks, and as we
shall demonstrate in section \ref{subs:tsmbresults}, there are indeed isolated
real eigenvalues and hence a sign problem for
the adjoint staggered model at large chemical potential $\mu$.
\subsection{Symmetry breaking pattern}
\label{subs:symmbreaking}
In the chiral limit, the action has a $\U{N}_L\otimes\U{N}_R$ symmetry,
which for staggered fermions is manifest as independent $\U{N}$
symmetries for the even and odd sites.  At $\mu=0$ this enlarges to a
$\U{2N}$ symmetry.  This can be seen most easily by introducing new
fields,
%
\begin{equation}
\bar{X}_e = (\chibar_e,-\chi^{\rm{tr}}_e\tau_2) \qquad X_o = \begin{pmatrix}
\chi_o\\-\tau_2\bar\chi_o^{\rm tr}
\end{pmatrix}
\end{equation}
for fundamental quarks, and
\begin{equation}
\bar{X}_e = (\chibar_e,\chi^{\rm{tr}}_e) \qquad X_o = \begin{pmatrix}
\chi_o\\\bar\chi_o^{\rm tr}
\end{pmatrix}
\end{equation}
%
for adjoint quarks.  The action can then be written as
\begin{equation} 
S={1\over2}\sum_{x \text{even}, \nu}\eta_\nu(x)\left(L_{+}(x)-L_{-}(x)\right)
\end{equation}
where
\begin{align}
L_{+}\!&=\! \bar X_e(x)\begin{pmatrix}e^{\mu\delta_{\nu,0}}& \\
              &\hspace{-6mm}e^{-\mu\delta_{\nu,0}} \end{pmatrix}
                           U_\nu(x)X_o(x\!\!+\!\!\hat\nu) \\
L_{-}\!&=\!\bar X_e(x)\begin{pmatrix}e^{-\mu\delta_{\nu,0}}&\\
              &\hspace{-6mm}e^{\mu\delta_{\nu,0}}\end{pmatrix}
U_\nu^\dagger(x\!\!-\!\!\hat\nu)X_o(x\!\!-\!\!\hat\nu)
\end{align}
In the continuum, the equivalent fields are
\begin{align}
\text{fundamental:} &\qquad 
\Psi = \begin{pmatrix}\psi_L\\ \sigma_2\tau_2\psi_R^{*}\end{pmatrix}
\\
\text{adjoint:} & \qquad
\Psi = \begin{pmatrix}\psi_L\\ \sigma_2\psi_R^{*}\end{pmatrix}
\end{align}
which gives the lagrangian
\begin{equation}
\mathcal{L} = i \Psi^\dagger \sigma_\nu(D_\nu-\mu B_\nu)\Psi
\end{equation}
where
\begin{equation}
B_\nu = B\delta_{\nu 0} \, ; \qquad 
B = \begin{pmatrix} 1 & 0 \\ 0 & -1 \end{pmatrix}
\end{equation}
The chiral condensate can be written in terms of the new fields,
\begin{equation}
\bar\chi\chi=
\bar X_e\!\begin{pmatrix}&\hspace{-3mm}\One\\\pm\One&\end{pmatrix}\!{T\over2}\bar X_e^{tr}+
X_o^{tr}\!\begin{pmatrix}&\hspace{-3mm}\One\\\pm\One&\end{pmatrix}\!{T\over2}X_o
\label{eq:condf}
\end{equation}
where the $+$ sign is for fundamental fermions and the $-$ sign for
adjoint, while $T$ is the unitary operator defined in section
\ref{subs:sign}.  A nonzero chiral condensate thereby breaks down the
$\U{2N}$ symmetry to O($2N$) for fundamental fermions and Sp($2N$) for
adjoint fermions, giving rise to $N(2N+1)$ and $N(2N-1)$ Goldstone modes
respectively.  Of these, there will be $N^2$ mesonic states, while the
remaining $N(N\pm1)$ will be diquarks.  From this we see that in the
case of $N=1$ adjoint fermions, and only in this case, are there no
diquark Goldstone modes.

For $m\not=0$, all states remain degenerate, gaining
masses $m_\pi\propto\sqrt{m}$ in accordance with standard PCAC arguments.
As the chemical potential $\mu$ increases from zero, a ground state
containing baryonic matter is promoted, signalled by a non-zero
value for the baryon number density 
\begin{equation}
\begin{split}
n={1\over2}\big\langle&\bar\chi(x)\eta_0(x)[e^\mu U_0(x)\chi(x+\hat0)
\\
& + e^{-\mu}U_0^\dagger(x-\hat0)\chi(x-\hat0)]\big\rangle. 
\end{split}
\label{eq:n}
\end{equation}
At zero temperature
$n$ thus serves as an order parameter for an {\em onset} phase transition
occuring at some $\mu_o$ separating the vacuum from a state containing matter.
A naive energetic argument would 
suggest that the onset transition should occur for a value of $\mu_o$ 
equal to 
the mass per baryon charge  of the lightest particle carrying 
non-zero baryon number. For the models discussed in the previous paragraph
in which some of the Goldstone modes are diquark states, those states will
be the lightest baryons in the spectrum. This means that for most
variants of Two Color QCD we expect
$\mu_o\simeq m_\pi/2$, in contrast to the much larger value 
$m_N/3$ expected in physical QCD.
The exception is $\text{SU(2)}^{N=1 \text{stagg}}_{\text{Adj}}$.

\section{ALGORITHMS}
\label{sec:algorithms}

We have studied Two Color lattice QCD with $N=1$ adjoint flavors of
staggered fermions, using two different simulation algorithms: the
hybrid Monte Carlo (HMC) algorithm \cite{DKPR}, and a Two-Step
Multi-Bosonic (TSMB) algorithm \cite{TWO_STEP}.  In \cite{us} we
described in detail how both algorithms are defined for the model
above. Here we only remind that, in order to optimise the
performances, in both cases we need to tune the simulation parameters.
In HMC we tuned the mean trajectory length and the size of the
discretised time step. In TSMB we have a larger set of parameters: the
interval $[\epsilon,\lambda]$ over which the polynomial approximation
is performed, the degrees of the four polynomials ($n_i$,
$i=1,\ldots,4$), and the number of Metropolis, heatbath and
overrelaxation iterations for the gauge and the boson fields
respectively.

The first observation is that
as the ratio $\mu / \sqrt{m}$ increases some eigenvalues of the fermionic 
matrix approch zero and any simulation becomes hard in that region.
By increasing $\mu / \sqrt{m}$ further, some eigenvalues get a negative
real part. It turns out that the model allows configurations with a
negative determinant (when an odd number of real negative eigenvalues
appear). At this point TSMB and HMC display different behaviours. TSMB
simulations can easily sample configurations with both signs of 
$\det{M}$, for any combination
of the simulation parameters that we used. HMC simulations (at least
for the combinations of trajectory lengths and time steps that we used, which
are dictated by efficiency considerations) were never able to change
the sign of $\det{M}$.

The reason for this difference must reside in the updating of the 
gauge fields, since
$\det{M}$ is only defined in term of these. In the case of HMC the 
accept/reject step is performed only after updating the whole configuration
and the exact action is used. In the case of TSMB the gauge field updating
is performed by a Metropolis algorithm and the action is given by
a polynomial approximation realised by means of auxiliary boson fields.
These conditions apparently allow TSMB to change the sign of $\det{M}$
more easily than HMC. The exactness of the TSMB algorithm is guaranteed by 
a final reweighting step, which is necessary since no polynomial 
approximation is sufficient when dealing with arbitrarily small eigenvalues.
\section{RESULTS}
\label{sec:results}
\subsection{Autocorrelation studies}
\label{subs:auto}
In order to establish the errors of our numerical studies we
analysed the autocorrelation times of both algorithms
and expressed them in the common unit of measure of 
{\em matrix multiplications}
(appropriately corrected with a factor that 
takes into account that TSMB spends more time in other kinds 
of operation). The number of matrix multiplications per sweep is a
function of the parameters of the simulation \cite{us}.
It turns out, in general, that the gluonic observables have a much longer autocorrelation
time than the fermionic ones, sometimes by even two orders of magnitude. 
In the following we will consider only the former, in order to be conservative. 
At zero density  we could determine the autocorrelation time 
with sufficient precision for both algorithms. The results are shown in table
\ref{autocorr_tab}.
\begin{table*}[th]
\begin{center}
\caption{\label{autocorr_tab}
 Integrated autocorrelation of the plaquette $\tau_{int}^{plaq}$ for
 TSMB and HMC at $\beta=2.0,\; m=0.1,\; \mu=0.0$ on a $4^3\times8$ lattice.
 For TSMB we use polynomial orders: $n_1,\; n_2=90,\; n_3=120$;
 interval of approximation: $[\epsilon=0.005,\lambda=10.0]$.
 For both algorithms $R_{acc}$ is the
 acceptance rate of the correction step, and $N_{sweep}$ is the
 length of runs in sweeps or trajectories.  
 $N_{mult}$ is the number of matrix multiplications per sweep.
 $N_{int}^{plaq}$ is the autocorrelation in number of matrix
multiplications.  
}
\end{center}
\begin{center}
\begin{tabular}{cccccrcl}
\hline
 Alg. & $n_1$  &  $R_{acc}$  &  $N_{sweep}$  &  $\tau_{int}^{plaq}$  & 
 $N_{mult}$  &  $N_{int}^{plaq}$  &  $\langle \Box\rangle$ \\
\hline

 TSMB & 24 & 0.68 & 283000 & 420(20) & 1500 & $6.30 \cdot 10^5$ & 0.5676(8) \\ 
 TSMB & 24 & 0.68 & 110000 & 430(35) &  500 & $2.15 \cdot 10^5$ & 0.5687(10) \\ 
 TSMB & 16 & 0.32 & 195000 & 360(30) &  460 & $1.66 \cdot 10^5$ & 0.5684(7) \\ 
\hline
 HMC & - & 0.83 & 35000 & 10  &  19500 & $1.9 \cdot 10^5$ & 0.5682(4) \\ 
\hline
\end{tabular}
\end{center}
\end{table*}
%
The runs in the table are sufficiently long for the measurement of the
integrated autocorrelations.
The run in the first line has relatively more gauge update sweeps
compared to the boson field updates: it has $N_M = 12$ Metropolis
gauge sweeps per $N_H =2$ heatbath and $N_O = 1$ overrelaxation bosonic
sweeps.
This is obviously not advantageous for the autocorrelation.
In the other two runs $N_M = 4$ which is substantially better.
The difference between the second and third lines is in the number of
auxiliary boson fields (equal to the rank of the first polynomial).
The table shows that a low acceptance about 30\% is somewhat better
than the higher one near 70\%.
Further optimisation of the choice of TSMB parameters in this point
is still possible but $N_{int}^{plaq}$ is already smaller than
the corresponding number in HMC.

Our longest TSMB run at $\mu=0.4$ is not long enough for an accurate
determination of the integrated plaquette autocorrelation.
After 80000 sweeps (on a $4^3 \times 8$ lattice) the obtained 
autocorrelation estimate is $\tau_{int}^{plaq} \simeq 2400 \;
(N_{int}^{plaq} \simeq 6\cdot10^6)$.
The obtained result is good enough for an order of magnitude
estimate of $\tau_{int}^{plaq}$ but the real value may be somewhat
larger. Also in the case of HMC, the longest run of 18000 trajectories
was not sufficient to determine the autocorrelation.

Since the point at $\mu=0.4$ appears so difficult, we started an
analysis of autocorrelation at point $\mu=0.36$. Our longest
run (40000 sweeps) is still too short to provide an accurate
determination. However we can at least estimate the order of 
magnitude of $\tau_{int}^{plaq} \simeq 500 \;(N_{int}^{plaq} \simeq 8\cdot10^5)$.

\subsection{Reweighting}
\label{subs:reweight}

\begin{figure}[tb]
\includegraphics[width=\colw]{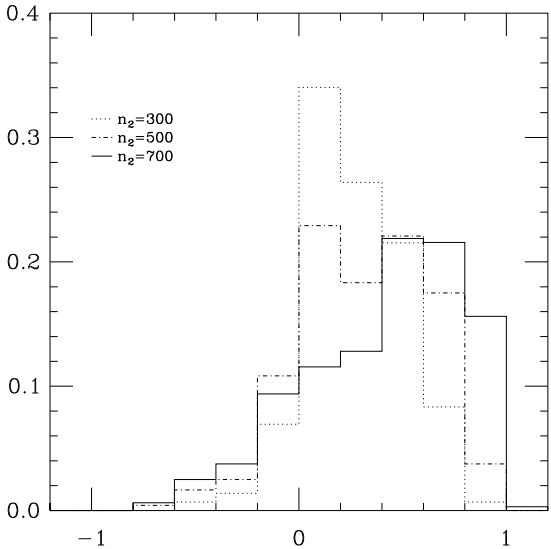}
\vspace{-0.7cm}
\caption{Examples of reweighting factors for different choices
of the second polynomial, at $\beta=2.0, m=0.1, \mu=0.36$ with $n_1=64$.
\label{fig:reweight}}
\end{figure}

 The final precision in the TSMB algorithm is achieved by reweighting
 the gauge configurations at the evaluation of expectation values.
 In a model with $\mbox{det}M$ possibly negative, as 
 considered here, the sign of the determinant is also taken into account
 in the reweighting.
 The choice of the order of the second polynomial $n_2$ has an important
 effect on the reweighting.
 For sufficiently large $n_2$ the two-step approximation can be so good
 that the effect of reweighting is negligible compared to the
 statistical errors.
 In fact, this ideal situation can be achieved in the low density phase
 of our model where negative determinants practically never occur.
 In the high density phase, however, negative determinants play an
 important r\^ole and it is more advantageous to have a non-negligible
 reweighting.
 (This is also required by the very small eigenvalues which can be treated
 exactly in the reweighting step.)
 Of course, a very low value of $n_2$ is not optimal either because then
 the typical reweighting factors become too small and the effective
 statistics is substantially reduced.
 The effect of choosing the order of the second polynomial on the
 distribution of reweighting factors is shown by fig. \ref{fig:reweight}.
%
%
%
%
\subsection{Physics results from HMC}
\label{subs:hmcresults}
We used three distinct quark masses on 
the $4^3\times8$ lattice at $\beta=2.0$, 
and explored values of $\mu$ up to and including 0.8
for $m=0.1$, $\mu=0.7$ for $m=0.05$, and $\mu=0.5$ for $m=0.01$. 

\begin{figure}[tb]
\includegraphics[width=\colw]{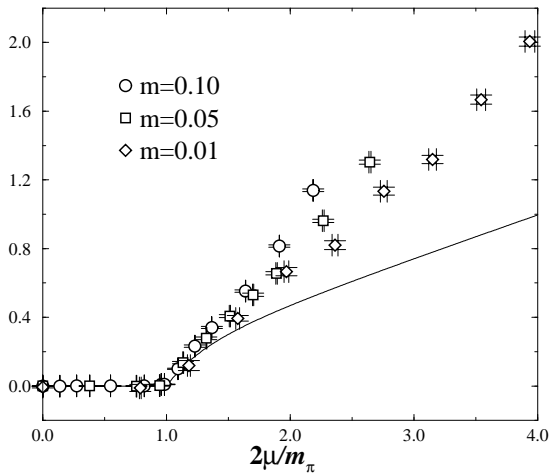}
\vspace{-0.8cm}
\caption{Baryon density vs. chemical potential using the rescaled variables
of eq. (\ref{eq:rescale}).
\label{fig:uniden}}
\end{figure}

Here we measured the chiral condensate $\langle\bar\chi\chi\rangle$ 
the baryon number density $n$ (\ref{eq:n}), the plaquette and
the pion mass $m_{\pi}$.
If we define the rescaled variables 
$x=2\mu/m_{\pi0}$, 
$y=\langle\bar\chi\chi\rangle/\langle\bar\chi\chi\rangle_0$, and
$\tilde n=m_{\pi0}n/8m\langle\bar\chi\chi\rangle_0$ (where the 0 subscript 
denotes values at zero chemical potential), then $\chi$PT predicts
(in the limit of small $m$ and $\mu$) that all data should
fall on the lines \cite{KSTVZ}
\begin{equation}
y=\begin{cases}
	1&\!x\!<\!1\\
	1\over x^2&\!x\!>\!1
  \end{cases}
\quad
\tilde n=\begin{cases}
	0&\!x\!<\!1\\
	{x\over4}\left(1-{1\over x^4}\right)
&\!x\!>\!1
	 \end{cases}
\label{eq:rescale}
\end{equation}
\begin{figure}[tb]
\includegraphics[width=\colw]{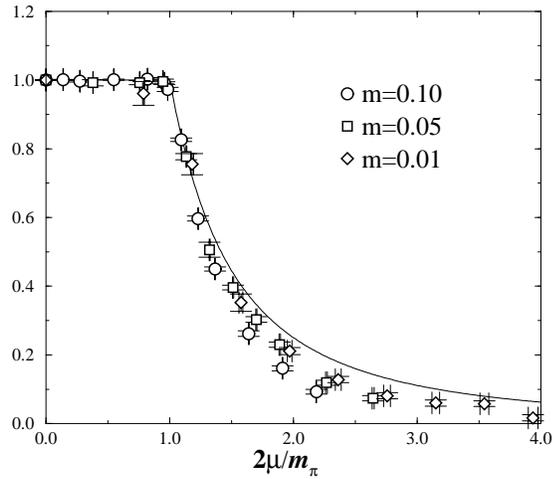}
\vspace{-0.8cm}
\caption{Chiral condensate vs. chemical potential
using the rescaled variables of eq. (\ref{eq:rescale}). 
\label{fig:unipbp}}
\end{figure}
In Figs~\ref{fig:unipbp} and \ref{fig:uniden} we show the rescaled
variables $y$ and $\tilde n$ respectively as functions of $x$.
\begin{figure}[tb]
\includegraphics[width=\colw]{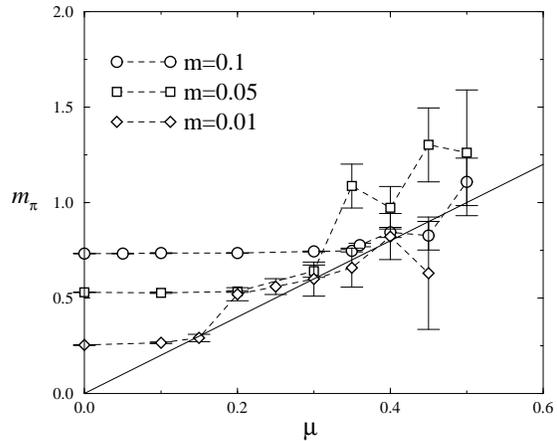}
\vspace{-0.8cm}
\caption{$m_\pi$ vs. $\mu$, for the three different quark masses. Also
shown is the line $m_\pi=2\mu$.
\label{fig:pion}}
\end{figure}
The data collapse very nicely onto a universal curve, corresponding
quite closely to the prediction (\ref{eq:rescale}). The systematic departures
from the theoretical curves for $x<2$, 
downwards for the condensate data and upwards
for the baryon density, may well be explicable by higher order corrections in 
$\chi$PT. Our more recent results from the regime $x>2$, however, suggest a
dispersion in the data from different $m$, and hence a possible 
breakdown of $\chi$PT at higher densities. This may be due to new thresholds
as particles other than Goldstones are induced into the ground state, or even a 
further phase
transition \cite{us}. 

Whilst the approximate quantitative agreement between our results and the 
theoretical predictions 
of \cite{KSTVZ} is gratifying, it also contradicts the symmetry-based arguments 
of section \ref{sec:symmetries} that there are no baryonic Goldstones for $N=1$
staggered flavor, and no gauge-invariant local diquark condensate. 
We believe that this is because the HMC simulations fail
to take account of the determinant sign (or indeed even to change it)
ie.\ that simulations with functional weight $\vert\det{M}\vert$
yield broadly similar results to those with weight $\det^2 M$. The 
premature onset at $\mu_o=m_\pi/2$
is therefore a direct manifestation of the sign and/or ergodicity problems.

In Fig.~\ref{fig:pion} we show results for $m_\pi$, obtained using a 
standard cosh fit to the meson correlator over all 8 timeslices. The fits are
quite stable in the low density phase, but the correlators become very noisy
once $n>0$, resulting in a much reduced precision. It is significant,
however, that our results in the dense phase are at least
consistent with the $\chi$PT
prediction $m_\pi=2\mu$ \cite{KSTVZ}.

\begin{figure}[tb]
\includegraphics[width=\colw]{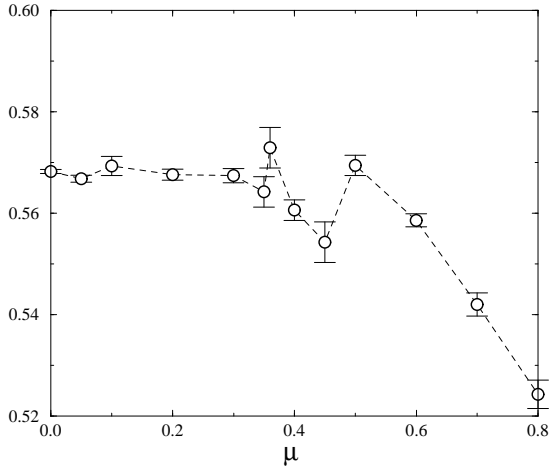}
\vspace{-0.9cm}
\caption{Average plaquette
vs. $\mu$ for $m=0.1$.
\label{fig:plaq}}
\end{figure}

Finally we turn to the effect of the chemical potential on the gauge fields.
Since this can only be communicated via fermion loops, any effect we see
can be ascribed with certainty to dynamical fermions.
Gluonic observables, however, are also much more 
prone to auto-correlations as described in section \ref{subs:auto}, 
particularly as the quark mass is reduced. Systematic changes with $\mu$
are therefore quite difficult to observe. In this initial HMC study we 
have only measured the average plaquette; the results are shown in 
Fig.~\ref{fig:plaq}.
The data for $m=0.1$ show the plaquette remaining roughly constant
for $\mu<\mu_o$, before beginning to decrease. The points for $m=0.05,0.01$
have been omitted for clarity, but reveal a similar picture.
We interpret 
it as follows: for temperature $T=0$, all values of $\mu<\mu_o$ are physically
equivalent corresponding to the same physical state, namely the vacuum.
We only expect an effect on gluonic observables in the presence of matter,
ie.\ for $\mu>\mu_o$. To the extent that the results are constant for
$\mu<\mu_o$ we can be confident that our simulation has an effective 
$T\simeq0$. The decrease in the plaquette for 
$\mu>\mu_o$ may be due to the decrease in the
number of virtual quark--anti-quark pairs which may form due to the Exclusion
Principle
--- an effect known as {\em Pauli blocking\/}. This results
in a decrease of screening via vacuum polarisation, and hence an effective
renormalisation of the gauge coupling $\beta$ and consequent decrease of 
the plaquette. In the large-$\mu$ limit the lattice should become 
saturated with one quark of each color per site, and the plaquette assume its
quenched value \cite{HM2}.
\subsection{Physics results from TSMB}
\label{subs:tsmbresults}
Each TSMB simulation is characterised by a vector $n_i$ specifying the 
polynomial orders at each stage, as described in \cite{TWO_STEP}.
For each configuration generated, a reweighting factor $r$ and the sign of 
$\det{M}$ must be determined. On the relatively small lattices considered
here, we have been able to compute $\det{M}$ directly using
standard numerical methods.
The expectation value of an observable $O$ is then determined by the 
ratio
\begin{equation}
\langle O\rangle={{\langle O\times r\times sign\rangle}\over
                  {\langle r\times sign\rangle}}.
\label{eq:Osign}
\end{equation}
Here we present results from runs performed on a $4^3 \times 8$
lattice with  $\beta=2.0$, $m=0.1$ and three values of $\mu$.  The
parameters are given in table~\ref{tab:tsmbparams}.
\begin{table}[tb]
\caption{Parameters for the TSMB simulations.  $N_{\text{cfg}}$ is an
estimate of the number of independent configurations, using the
estimates of the autocorrelation time from section
\ref{subs:auto}.}
\label{tab:tsmbparams}
\begin{tabular}{dcrr}
\cc{$\mu$} & $(n_1,n_2)$ & $N_{\text{cfg}}$
 & $\langle r\times sign\rangle$ \\ \hline
0.0  & (16,100) &  90 \\
0.36 & (64,500) & 120 & 0.313(20) \\
     & (64,700) & 160 & 0.417(21) \\
0.4  & (100,1000) & 128 & 0.030(59) \\ \hline
\end{tabular}
\end{table}
Note that the polynomial orders required increase
with $\mu$. The point at $\mu=0$
was chosen to enable the TSMB algorithm to be tested against HMC, since 
both should yield identical results. The $\mu\not=0$ points were
chosen so as to have one value just past the HMC onset transition,
where the edge of the eigenvalue distribution just overlaps the line
$\mbox{Re}\lambda=0$ and a small percentage of negative determinant
configurations are expected, so that hopefully the sign problem is not too
severe, and one value fairly deep in the high density 
phase. 
 
Our results for the standard observables, together 
with the corresponding HMC results, are summarised in Table
\ref{tab:tsmb}. For TSMB at $\mu\not=0$ we also include 
observables determined separately in each sign sector, defined by
$\langle O\rangle_\pm=\langle O\times r\rangle_\pm/\langle r\rangle_\pm$.
\begin{table*}[tbh]
\caption{A comparison of results between TSMB and HMC. 
At $\mu=0.0$ reweighting was not necessary.  Due to the long
autocorrelation times, the errors in the HMC results, especially for
the plaquette, are probably underestimated.\label{tab:tsmb}}
\begin{tabular}{lddddd} \hline
 & \cc{$\mu$} & \multicolumn{3}{c}{TSMB} & \cc{HMC} \\ \cline{3-5}
 & & \cc{$\langle O\rangle$} & \cc{$\langle O\rangle_+$} & \cc{$\langle O\rangle_-$} & 
\\ \hline
$\langle\bar\chi\chi\rangle$ & 0.0 & 1.525(3) & & & 1.526(1) \\
 & 0.36 & 1.551(10) & 1.521(8) & 1.176(37) & 1.485(9) \\
 & 0.4  & 2.49(261) & 1.31(4) & 1.19(4) & 1.253(10) \\ \cline{2-6}

$n$ & 0.0 & 0.0000(13) & & & -0.0002(3) \\
 & 0.36 & -0.0003(80) & 0.0199(64) & 0.252(32) & 0.0172(28) \\
 & 0.4  & -0.65(179) & 0.14(3) & 0.23(3) & 0.1667(90) \\ \cline{2-6}

$\Box$ & 0.0 & 0.5681(8) & & & 0.5682(4) \\
 & 0.36 & 0.5607(9) & 0.5605(8) & 0.5580(16) & 0.5729(40)\\
 & 0.4 & 0.635(149) & 0.5656(16) & 0.5584(17) & 0.5612(30) \\ \cline{2-6}

$m_\pi$ & 0.0 & 0.7321(10) & & & 0.7327(4) \\
 & 0.36 & 0.768(35) & 0.798(23) & 1.40(71) & 0.7778(88) \\
 & 0.4  & 2.14(374) & 1.30(40) & 0.95(19) & 0.8712(34) \\ \hline
\end{tabular}
\end{table*}
We note that the agreement between HMC and TSMB at $\mu=0.0$ is 
good, as it should be. 
The results at $\mu=0.4$ have too large errors and we cannot
derive any conclusion. The results at $\mu=0.36$ show 
an acceptable agreement between HMC and the positive sector of TSMB. 
However, the results for the negative determinant sector are
significantly different.  (A similar trend may be seen at $\mu=0.4$,
but here the difference is still within $2\sigma$, and thus not
statistically significant.)  The effect of the sign on the total
average --- especially for the baryon density --- is quite
interesting.  Even if the errors are
still large, we can see that $n$ is already definitely non
zero in the positive sector. The inclusion of the negative sector
has the effect of bringing back the average $n$ to zero --- although
it should be pointed out that the TSMB average is also still only
about $2\sigma$ away from the HMC result.
This mechanism is non-trivial since at that point there are
very few configurations with negative determinant.
For the chiral condensate, the effect of the sign is qualitatively
similar in that the inclusion of the sign counteracts the suppression
observed in HMC --- but we are unable at this stage to draw any
quantitative conclusions.

These observations suggest that at this value of
$\mu$ the system is still in the low density phase, and hence 
$\mu_{\text{o TRUE}}>\mu_{\text{o HMC}}$. This is
in accord with our symmetry-based
arguments that for $N=1$ flavors of adjoint staggered fermion there are no
baryonic Goldstones and hence no early onset. 
\section{CONCLUSIONS}
We presented our progress in the numerical study of
$\text{SU(2)}^{N=1\, \text{stagg}}_{\text{Adj}}$. We
have seen that when the model is constrained to the sector
with positive determinant it displays a good agreement
with the expectations of $\chi$PT. The departure from 
$\chi$PT at large density is now more clear and deserves
further investigation. We also confirmed, with narrower
errorbars, the presence of a delayed onset when
the full model is taken into account. We learnt how
to perform a numerical analysis of this difficult
model through the use of a TSMB algorithm which is well
suited to sample correctly the sign. 
In future  we shall also study the di-quark condensates 
suggested in \cite{us} in order to understand fully
the effect of the sign in the region of the parameters
where this is possible.
\section*{Acknowledgements}
This work is supported by  the TMR-network ``Finite temperature phase 
transitions in particle physics'' EU-contract ERBFMRX-CT97-0122.
LS thanks the Royal Society for the benefit of a conference grant.
Numerical work was performed using a Cray T3-E at NIC,
J\"ulich and an SGI Origin2000 in Swansea.


\begin{thebibliography}{99}
%
%
\bibitem{ARW} M. Alford, K. Rajagopal and F. Wilczek, Phys. Lett. {\bf B422}
(1998) 247, Nucl. Phys. {\bf B\,537} (1999) 443;\\
R. Rapp, T. Sch\"afer, E.V. Shuryak and M. Velkovsky, Phys. Rev. Lett. {\bf81}
(1998) 53.
%
\bibitem{Cha} S. Chandrasekharan, these proceedings.
%
\bibitem{KSTVZ} J. Kogut,
M.A. Stephanov, D. Toublan, J.J.M. Verbaarschot
and A. Zhitnitsky,
Nucl.~Phys.~{\bf B\,582}, 477 (2000)
%
\bibitem{HM2} S.J. Hands and S.E. Morrison, talk at {\sl Understanding
Deconfinement in QCD\/},
eds. D. Blaschke
{\it et al\/}, p. 31, {\tt hep-lat/9905021}.
%
\bibitem{us} S.~Hands {\em et al\/},
Eur.\ Phys.\ J.~{\bf C}, DOI 10.1007/s100520000477;
{\tt hep-lat/0006018.}
%
\bibitem{DKPR} S. Duane {\em et al\/},
Phys. Lett. {\bf B\,195} (1987) 216.
%
\bibitem{TWO_STEP}
I. Montvay,
Nucl.\ Phys.\ {\bf B\,466} (1996) 259.
%

\end{thebibliography}
\end{document}